\documentclass[12pt,a4paper]{article}

\usepackage{cite,graphicx,caption,subcaption}
\usepackage{amsfonts,amsmath,amssymb}
\usepackage{braket,slashed,mathrsfs,tensor}
\usepackage{bbm,multicol,placeins,wrapfig}
\usepackage{array,multirow}
\usepackage{epstopdf,dsfont}

\begin{document}

\begin{center}
{\Large\bf Current Status of the Standard Model Prediction for the\\[2mm]
$B_s \to \mu^+ \mu^-$ Branching Ratio\footnote{
Contribution to the {\it Symmetry} Special Issue ``Symmetries and Anomalies in Flavour Physics''.}}\\[6mm]
{\bf Mateusz Czaja and Miko{\l}aj Misiak}\\[4mm] 
{\it Institute of Theoretical Physics, Faculty of Physics, University of Warsaw, Poland.}
\end{center}

\ \\
{\bf Abstract:} The rare decay $B_s \to \mu^+ \mu^-$ provides an important constraint on
possible deviations from the Standard Model in $b$-$s$-$\ell$-$\ell$
interactions. The present weighted average of its branching ratio
measurements amounts to $(3.34 \pm 0.27)\times 10^{-9}$, which remains in
good agreement with the theoretical prediction of $(3.64 \pm 0.12)\times
10^{-9}$ within the Standard Model. In the present paper, we review
calculations that have contributed to this prediction, and discuss the
associated uncertainties.

\section{Introduction}

The Higgs boson discovery~\cite{ATLAS:2012yve,CMS:2012qbp}
through direct production at the LHC completed the experimental search
for the Standard Model (SM) particle content. Since then,
no clear signal for Beyond-Standard Model (BSM) particle production
has been seen at the high-energy frontier of experimental
particle physics. Consequently, more and more focus is being
shifted towards precise studies of rare processes that
are sensitive to corrections from BSMs.
  
In particular, decays of the $B$ meson mediated through
Flavour-Changing Neutral Currents (FCNCs) have been a very active area
of research. Theoretical studies of the $B$ mesons are
greatly aided by the framework of Heavy Quark Expansion
which, in many cases, allows us to parameterize effects
of their hadronic structure through a series of non-perturbative
matrix elements suppressed by powers of $\Lambda_{QCD}/M_B$.
Moreover, while the FCNC-mediated processes are loop-suppressed
in the SM, they can receive sizeable tree-level contributions from
BSMs, which underlines them as primary candidates for
observables where indirect signals from new particles may
be detected~\cite{Altmannshofer:2021qrr}. The rich phenomenology
of rare $B$ meson decays is being actively studied by the
LHC experiments and Belle~II that follow and extend past
investigations at CLEO, Belle and BABAR. To fully take advantage of
the current and future experimental data, improvements in
precision of the SM predictions are often necessary.

\begin{figure}[t]
\begin{center}
\includegraphics[width=10.5 cm]{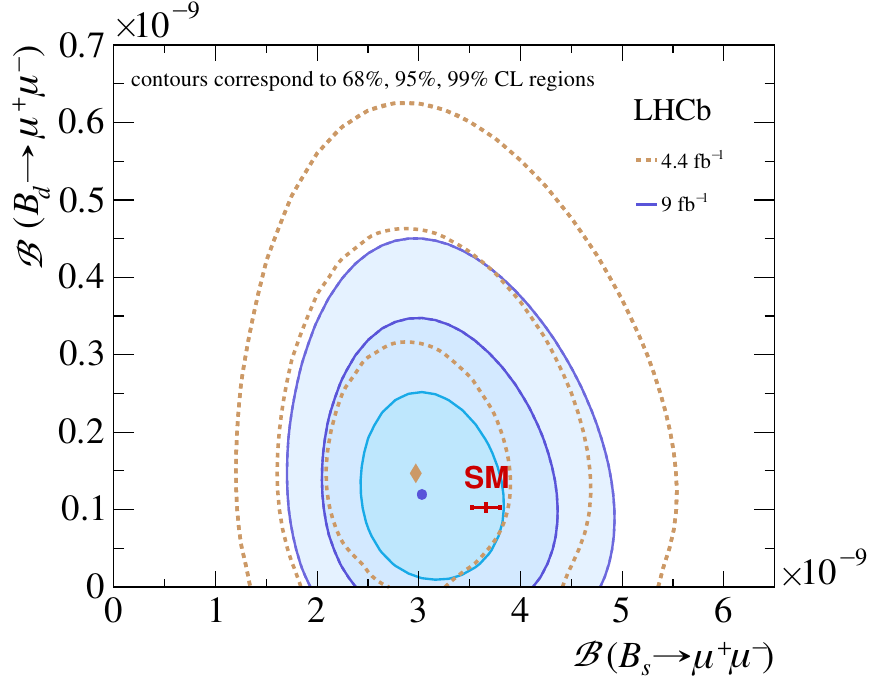}
\caption{LHCb measurements of $\overline{\mathcal B}_{s\mu}$ and $\overline{\mathcal B}_{d\mu}$
at 4.4 and 9$\,{\rm fb}^{-1}$ integrated luminosity, compared to
the current SM predictions. Plot from Fig.~2 of
Ref.~\cite{Bsmm-LHCb-2022}.\label{fig1}}
\end{center}
\end{figure}

One of the most interesting rare decays of the $B$ meson and the focus
of this review is the $B_s\to\mu^+\mu^-$ channel that was first
observed over a decade ago~\cite{FirstBsmm}. Since then, the
experimental precision for its average time-integrated
branching ratio $\overline{\mathcal B}_{s\mu}$ has reached
$\mathcal{O}(10\%)$~\cite{CMS:2019bbr,CMS:2022mgd,Bsmm-LHCb-2022,ATLAS:2018cur}.
The current world average reads~\cite{PDG2024}
%
%
\begin{equation}
\label{experiment}
\overline{\mathcal B}_{s\mu}^{\;\rm exp} =(3.34 \pm 0.27)\times10^{-9}.
\end{equation}
The similar $B_d\to\mu^+\mu^-$ channel is suppressed by a factor
$|V_{td}/V_{ts}|^2\approx0.04$, which has a negative effect on
experimental precision due to significantly reduced statistics, as
illustrated in Fig.~\ref{fig1}.

The SM description of $B_s\rightarrow\mu^+\mu^-$ is greatly simplified
by a factorization of long- and short-distance Quantum Chromodynamics
(QCD) effects. In contrast to many other $B$ decays, the SM decay
amplitude of $B_s\rightarrow\mu^+\mu^-$ depends, to a very good
approximation, only on a single non-perturbative hadronic quantity,
namely the $B_s$-meson decay constant $f_{B_s}$. Its determinations from lattice
simulations (see below) are mature and precise. Moreover, the hard QCD
and electroweak (EW) corrections are, to a very good approximation,
contained within a single Wilson coefficient $C_A$, which can be
calculated in a standard, perturbative matching procedure as a series
in the strong and electromagnetic couplings $\alpha_s$ and
$\alpha_e$. Thanks to these properties, the SM calculations have
reached a few percent accuracy. The current SM prediction for the
branching ratio reads
\begin{equation} \label{brsm}
\overline{\mathcal B}_{s\mu}^{SM}=(3.64\pm0.12)\times10^{-9}.
\end{equation}
We describe its evaluation in the next sections. The numerical value
in Eq.~(\ref{brsm}) has been obtained by updating the input parameters
in the semi-numerical expressions of Ref.~\cite{Bobeth:2013uxa}, and
including the power-enhanced QED correction from
Refs.~\cite{Beneke:2017vpq,Beneke:2019slt} that amounts to around
$-0.5\%$.  A difference with respect to
$(3.66\pm0.14)\times10^{-9}$ in Ref.~\cite{Beneke:2019slt} is due to
the parameter update. The above SM prediction is well in agreement
with each individual measurement~\cite{CMS:2019bbr,CMS:2022mgd,Bsmm-LHCb-2022,ATLAS:2018cur},
as well as with their average in Eq.~(\ref{experiment}).

In the SM, on top of the aforementioned FCNC loop suppression, 
$\overline{\mathcal B}_{s\mu}$ receives a helicity suppression by the
mass-squared ratio $m_\mu^2/m^2_{B_s}$. One or both of these
suppressions may be lifted in models with additional Higgs doublets or
with a $Z'$. In effect, one finds restrictions on allowed parameter
spaces in, among others, Two Higgs Doublet Models
(2HDMs)~\cite{Arbey:2017gmh} and the Minimal Supersymmetric Standard
Model~\cite{Arbey:2011aa}. In addition, several time-dependent
observables in $B_s\rightarrow\mu^+\mu^-$ can be used to study
potential BSM ${\rm CP}$-violation mechanisms~\cite{Buras:2013uqa}.

The present review is organized as follows. In
section~\ref{section:Lag+br}, we present the effective
Lagrangian and the branching ratio formula for
$B_s\rightarrow\mu^+\mu^-$ in the SM. Sections~\ref{section:QCD}
and~\ref{section:EW} are devoted to describing,
respectively, the perturbative QCD and EW corrections to the
Wilson coefficient $C_A$. At the end of Section~\ref{section:EW},
the power-enhanced QED correction to $\overline{\mathcal B}_{s\mu}$ is
discussed. In section~\ref{section:num}, we summarize the current
parameter update and evaluate the SM prediction for
$\overline{\mathcal B}_{s\mu}$~(\ref{brsm}) together with the corresponding
uncertainty. We conclude in section \ref{section:conc}.  In~\ref{section:Appendix A},
we recall the derivation of the branching ratio formula. In~\ref{section:Appendix B},
we present the current SM prediction for $\overline{\mathcal B}_{d\mu}$,
i.e.\ the average time-integrated branching ratio of $B^0 \to \mu^+\mu^-$.

\section{The effective Lagrangian and the branching ratio formula}\label{section:Lag+br}

The effective Lagrangian used to describe $B_s\rightarrow l^+l^-,$
$l\in\{e,\mu,\tau\}$ in the SM is obtained through
simultaneous integrating out of all fields heavier than the $b$
quark at the scale $\mu_0=\mathcal{O}(m_t)$. It has the form
\begin{equation} \label{main lag}
\mathcal{L}=\mathcal{L}_{\text{QCD}\times\text{QED}}(\text{fields lighter than }W) + \left[ N\sum_nC_nQ_n+\mathrm{h.c.} \right],
\end{equation}
where
\begin{equation} \label{N def}
N\equiv\frac{V_{tb}^*V_{ts}G_F^2M_W^2}{\pi^2},
\end{equation}
$V_{ij}$ are the Cabibbo-Kobayashi-Maskawa (CKM) matrix elements,
$G_F$ is the Fermi constant, while $M_W$ is the $W$-boson mass defined
in the on-shell renormalization scheme. The local operators $Q_n$ are
polynomials in the light fields and their derivatives.  The
Wilson coefficients $C_n$ can be treated as real-valued (up to
negligible corrections) once the global normalization factor $N$ is
set as in Eq.~(\ref{N def}). The operators $Q_n$ are of mass-dimension
5 or higher, and have to be suppressed by powers of $1/M_W$ necessary
to make the overall mass-dimension of $\mathcal{L}$ equal to 4. At the
leading order in $1/M_W$ and $\alpha_{e}$, it is sufficient to
consider the operators $Q_n$ where a
$\Delta B=-\Delta S=-1$ flavour-changing quark current
multiplies a lepton current. Moreover, the quark current must violate
parity to annihilate the pseudoscalar $B_s$ meson. Once the Lorentz
invariance is imposed, one is left with the following four operators
only:
\begin{equation}
\label{4operators}
\begin{split}
Q_A&\equiv[\bar{l}\gamma_\alpha\gamma_5l][\bar{b}\gamma^\alpha\gamma_5s]\equiv[\bar{l}\gamma_\alpha\gamma_5l] j^\alpha_A~~,\\
Q_V & \equiv [\bar{l}\gamma_\alpha l] j^\alpha_A~~,\\
Q_P&\equiv[\bar{l}\gamma_5l][\bar{b}\gamma_5s]\equiv[\bar{l}\gamma_5l]j_P~~,\\
Q_S&\equiv[\bar{l}l]j_P~~.
\end{split}
\end{equation}

The Lagrangian~(\ref{main lag}) can be used to derive the formula for
$\overline{\mathcal B}_{s\mu}$. A sketch of the derivation is given
in~\ref{section:Appendix A}. While evaluating the contribution of
$Q_A$ there, it becomes clear that $Q_V$ does not contribute at the
leading order in $\alpha_{e}$. As far as $Q_P$ and $Q_S$ are
concerned, their Wilson coefficients are computed in a matching to
full SM amplitudes with the quark and lepton currents
exchanging Higgs bosons. Such contributions are suppressed by
$m_b^2/M_W^2$, and can be neglected as being of the same order as
dimension-8 operator effects. Hence, neglecting these tiny
effects, $\overline{\mathcal B}_{s\mu}$ in the SM depends on $C_A$ only. The
explicit expression reads
\begin{equation}
\label{simplifiedbr}
\overline{\mathcal B}_{s\mu}^{SM}
=\frac{|N|^2M_{B_s}^3f_{B_s}^2}{8\pi\Gamma_H^s}\beta r^2|C_A|^2+\mathcal{O}\left(\alpha_{e},\frac{m_b^2}{M_W^2}\right),
\end{equation}
where $r\equiv 2m_\mu/M_{B_s}$, $\beta\equiv\sqrt{1-r^2}$, while
$\Gamma_H^s$ is the heavier mass eigenstate width in the
$B_s$--$\overline{B}_s$ system. Finally, the $B_s$ decay constant $f_{B_s}$ is defined through the relation
\begin{equation}
\label{fdef}
\braket{0|j^\alpha_A(x)|B_s(p)}\equiv i p^\alpha f_{B_s} e^{-ipx}.
\end{equation}
It is calculated using lattice QCD methods with errors at a
sub-percent level. The current world average based on $2+1+1$
simulations~\cite{Bazavov:2017lyh,ETM:2016nbo,Dowdall:2013tga,Hughes:2017spc}
alone amounts to~\cite{FlavourLatticeAveragingGroupFLAG:2021npn}
\begin{equation}
f_{B_s}=(230.3\pm1.3)\,\text{MeV}.
\end{equation}

In several popular BSMs, the Wilson coefficients $C_S$ and
$C_P$ can become comparable in size to $rC_A$. For example, in
the 2HDM-II with large $\tan\beta_H\equiv
v_2/v_1$ (the ratio of the vacuum expectation values of the two
Higgs doublets), one finds~\cite{Logan:2000iv},
\begin{equation}
C_S \simeq C_P \simeq \frac{\ln\rho}{\rho-1}\frac{m_\mu m_b}{4M^2_W}\tan^2\beta_H,
\end{equation}
where $\rho\equiv M_{H^+}^2/m_t^2$. The SM suppression factor
of $m_b^2/M_W^2$ can get compensated by values of
$\tan\beta_H=\mathcal{O}(50)$ or larger. The branching ratio
formula in the 2HDM becomes (see~\ref{section:Appendix A})
\begin{equation} \label{BR2HDM}
\overline{\mathcal B}^{2HDM}_{s\mu}=\frac{|N|^2M_{B_s}^3f_{B_s}^2}{8\pi\Gamma_H^s}\beta
\left[|rC_A-uC_P|^2+\frac{\Gamma_H^s}{\Gamma_L^s}|u\beta C_S|^2\right]+\mathcal{O}(\alpha_{e}),
\end{equation}
where $u\equiv M_{B_s}/(m_b+m_s)$ and $\Gamma_L^s$ is the
lighter mass eigenstate width in the $B_s$--$\overline{B}_s$
system.

Coming back to the SM expression for $\overline{\mathcal B}_{s\mu}$ in
Eq.~(\ref{simplifiedbr}), several comments concerning the
$\mathcal{O}(\alpha_e)$ terms there are necessary. First, all the
corrections of this order to $C_A$ are already known and included in
the numerical result in Eq.~(\ref{brsm}). They will be discussed in
subsection~\ref{subsection:EW1}. Some of them get enhanced
by $1/\sin^2\theta_W$, $m_t^2/M_W^2$ or $\ln^2M_W^2/m_b^2$, where
$\theta_W$ is the weak mixing angle. Second, some of the remaining
electromagnetic corrections in the $\mathcal{O}(\alpha_e)$ term in
Eq.~(\ref{simplifiedbr}) receive a power enhancement
by $M_{B_s}/\Lambda_{\text{QCD}}$~\cite{Beneke:2017vpq}.  They
come from virtual photons emitted by the $s$ quark,
and absorbed by the leptons. We shall comment on them in
subsection~\ref{subsection:EW2}. Their evaluation requires extending
the operator basis to include effects of other dimension-6
operators, not present in Eq.~(\ref{4operators}), such as
\begin{equation}
\label{q2def}
Q_2^\dagger \equiv [\bar{b}\gamma_\alpha P_Lc][\bar{c}\gamma^\alpha P_Ls]
\hspace{1cm} \mbox{or} \hspace{1cm}
Q_7^\dagger \equiv \frac{em_b}{16\pi^2} [\bar{b}\sigma^{\alpha\beta} P_L s] F_{\alpha\beta}~~ . 
\end{equation}
Third, the dependence of $C_A$ on the renormalization scale
$\mu_b$ (that arises due to QED effects only) induces around
$\mathcal{O}(0.3\%)$ uncertainty in
$\overline{\mathcal B}_{s\mu}^{SM}$~\cite{Bobeth:2013uxa}. Such a
dependence must be compensated by the yet unknown
$\mathcal{O}(\alpha_e)$ corrections that receive no extra
enhancement factors.

\section{The QCD corrections to $C_A$}\label{section:QCD}

As described in the previous section, at the leading order in
$\alpha_e$ and $m_b^2/M_W^2$, the only significant short-distance
parameter entering the SM branching ratio formula is the $C_A(\mu_b)$
Wilson coefficient, where we specifically indicate its dependence on
the low-energy renormalization scale $\mu_b=\mathcal{O}(m_b)$. It is
evaluated by demanding equality of corresponding Green's functions in
the SM and the effective theory~(\ref{main lag}) at the
renormalization scale $\mu_0=\mathcal{O}(m_t)$ at which the
electroweak degrees of freedom are decoupled. Computationally, the
simplest choice of a Green's function for the extraction of $C_A$ is
$G[\bar{b},s,\bar{l},l]$ with vanishing external momenta:
\begin{equation}
\label{matching eq}
G_{SM}[\bar{b},s,\bar{l},l](\mu_0)\rvert_{p_i=0}=G_{\rm eff}[\bar{b},s,\bar{l},l](\mu_0)\rvert_{p_i=0}.
\end{equation}
At the matching scale $\mu_0$, QCD on both sides is treated
perturbatively. This equation is then solved for $C_A(\mu_0)$
order-by-order in $\alpha_{s}$ and $\alpha_{e}$, resulting in a double
series
\begin{equation}
C_A(\mu_0)=\sum\limits_{m,n=0}^\infty\tilde{\alpha}_{s}^{m}\tilde{\alpha}_{e}^{n}C_A^{(m,n)}(\mu_0),
\end{equation}
where $\alpha_i(\mu_0)\equiv4\pi\tilde{\alpha}_i$ are the running
couplings renormalized in the $\overline{\text{MS}}$ scheme
at the scale $\mu_0$. In this section, we focus on the leading
terms in $\alpha_e$, namely $C_A^{(m,0)}$.

As we work at the leading order in $1/M_W$, all light masses in
the matching equation~(\ref{matching eq}) can be set to 0. In
dimensional regularization, all scaleless loop integrals 
vanish. In consequence, on the effective theory side, one is left
only with tree diagrams, including the UV-counterterm ones. On
the SM side, partially massive tadpoles have to be
calculated. Removing light masses from propagators in the SM Green's
function leads to spurious infrared divergences in loop integrals 
evaluated in $d=4-2\epsilon$ dimensions. The resulting additional
$\epsilon$-poles are not removed by renormalization constants of the
SM. Instead, they cancel in the matching equation~(\ref{matching
eq}) against the tree-level UV counterterms in the effective
theory.

\begin{figure}[t]
\begin{center}
\includegraphics[width=10.5 cm]{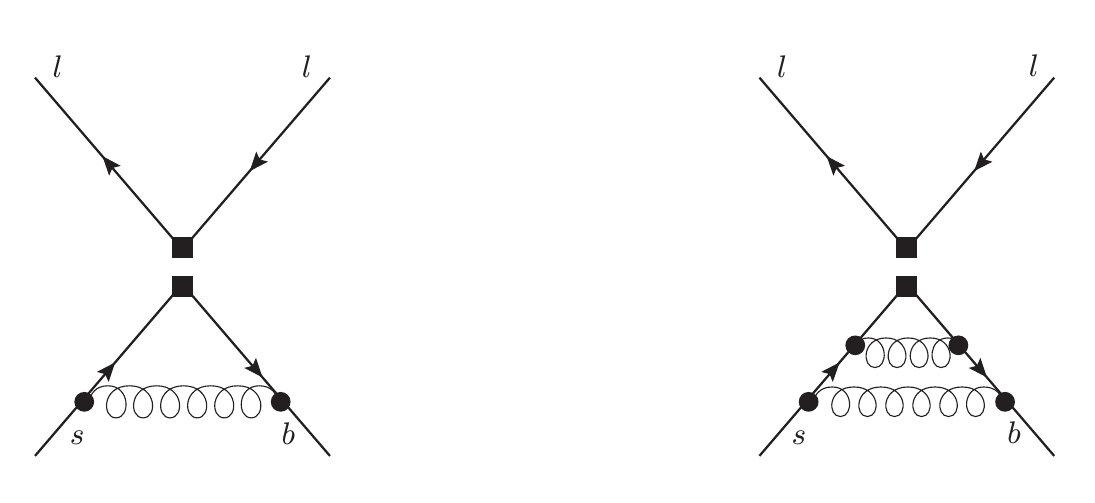}
\caption{Examples of the Feynman diagrams appearing in the calculation
of $Z_{ij}$. The double square in the middle denotes an insertion of
either $Q_A$ or $Q_A^E$. Diagrams from Fig.~2
of Ref.~\cite{Hermann:2013kca}.\label{fig2}}
\end{center}
\end{figure}

Once such a procedure is applied in our case, one has to
supplement the effective Lagrangian with additional operators
that vanish when $\epsilon\to 0$ due to the Dirac algebra
identities. Such operators are called evanescent. The UV-counterterms
with these operators cancel against some of the spurious
infrared divergence effects on the SM side.  Moreover, their Wilson coefficients
evaluated at lower orders affect the physical operator Wilson
coefficients at higher orders. For the $C_A^{(m,0)}$ terms, it is
sufficient to introduce only one evanescent
operator~\cite{Misiak:1999yg}:
\begin{equation}
Q_A^E=[\bar{b}\gamma_\alpha\gamma_\beta\gamma_\delta s][\bar{l} \gamma^\delta\gamma^\beta\gamma^\alpha l]-4Q_A.
\end{equation}
The bare fields, couplings and Wilson coefficients on the
effective side are replaced by the renormalized ones, with mixing 
occurring between the physical and evanescent operators:
\begin{equation}
\label{bareC}
C^{(b)}_AQ^{(b)}_A+C^{E(b)}_AQ^{E(b)}_A=Z_qZ_l(C_AZ_{NN}Q_A+C_AZ_{NE}Q_A^{E}+C_A^EZ_{EN}Q_A+C_A^{E}Z_{EE}Q_A^{E}),
\end{equation}
where $Z_q$ and $Z_l$ are the quark- and lepton-field
$\overline{\text{MS}}$ renormalization constants, respectively, with
$Z_l=1+\mathcal{O}(\alpha_e)$. To fix the renormalization constants
$Z_{ij}$, one demands that in the renormalized Green's
functions, the terms proportional to $C_A$ are finite when
$\epsilon\to 0$, while those proportional to $C_A^E$ vanish in this
limit~\cite{Dugan:1990df}. Examples of diagrams contributing to
$Z_{ij}$ at $\alpha_s$ and $\alpha_s^2$ are shown in
Fig.~\ref{fig2}. Results for the relevant $Z_{ij}$ up to
$\mathcal{O}(\alpha_e^0 \alpha_s^2)$ are given in Eq.~(15) of
Ref.~\cite{Hermann:2013kca}.

It is important to emphasize that $Z_{NN}=1$ to all orders in
QCD, once higher-dimensional operators and $\mathcal{O}(\alpha_e)$
effects are neglected. Therefore, the Renormalization Group
Equation (RGE) for $C_A$ is trivial at this level:
\begin{equation}
\mu\frac{d}{d\mu}C_A = \mathcal{O}(\alpha_e).
\end{equation}
Consequently, there is no RG evolution of $C_A(\mu)$ at the leading order in $\alpha_e$:
\begin{equation}
C_A(\mu_b)=C_A(\mu_0)+\mathcal{O}(\alpha_e).
\end{equation}

\begin{figure}[t]
\begin{center}
\includegraphics[width=15.5 cm]{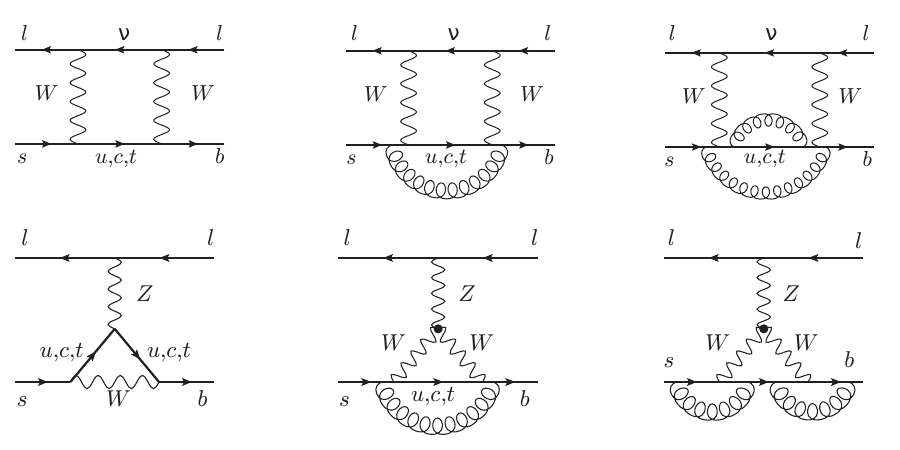}
\caption{Examples of diagrams entering the SM side of the
matching equation~(\ref{matching eq}) at various orders in
QCD. The $W$-boxes and $Z$-penguins are shown in the top
and bottom rows, respectively. Contributions of the orders
$\alpha_s^0$, $\alpha_s^1$, and $\alpha_s^2$ are arranged from left
to right. Diagrams from Figs.~1 and 4 of
Ref.~\cite{Hermann:2013kca}.\label{fig3}}
\end{center}
\end{figure}

The SM Green's function in Eq.~(\ref{matching eq}) receives QCD
corrections from two classes of diagrams: $W$-boxes and $Z$-penguins,
shown in Fig.~\ref{fig3}.  Such bare diagrams get renormalized using
lower-loop SM diagrams with counterterms. The QCD coupling constant
renormalization $Z_g$ has to be modified by an appropriate threshold
correction to match the coupling constant of the effective theory with
only 5 active flavours (see section 3 in
Ref.~\cite{Steinhauser:2002rq}). Similarly, the light-field
wave-function renormalization constants have to be shifted to account
for the decoupling threshold (see section 4 in
Ref.~\cite{Misiak:2004ew}). Heavy fields and masses are renormalized
in the $\overline{\text{MS}}$ scheme.

Once both sides of the matching equation~(\ref{matching eq}) have been
properly renormalized, the value of $C_A(\mu_0)$ can be extracted. All
the Dirac structures appearing on the SM side are mapped onto either $Q_A$
or $Q_A^E$, and their coefficients compared to the ones on the
effective side.

The procedure described in this section was completed in
Ref.~\cite{10.1143/PTP.65.297} at the leading order, followed by
Refs.~\cite{BUCHALLA1993225,BUCHALLA1993285,Misiak:1999yg} and
\cite{Hermann:2013kca} for the $\mathcal{O}(\alpha_s^1)$ and
$\mathcal{O}(\alpha_s^2)$ corrections, respectively. In the latter
case, high-order expansions in $y\equiv M_W/m_t$ and $w\equiv
1-M_W^2/m_t^2$ were computed, and their combination was used to obtain
a numerical result at the physical value of $M_W/m_t$.

\section{The electroweak corrections}\label{section:EW}

\subsection{The $C_A^{(0,1)}$ correction}\label{subsection:EW1}

The next-to-leading order EW correction $C_A^{(0,1)}(\mu_0)$ was first
obtained in Ref.~\cite{Bobeth:2013tba}.\footnote{
Let us note that it was called $c_{10}^{(2,2)}$ in that paper due to different notational conventions.}
Its calculation follows a similar pattern to the one described in the previous section. The main
difference comes from additional Dirac structures appearing in the
$\mathcal{O}(\alpha_e)$ corrections on the SM side of the
matching equation, which necessitate including additional operators in
the effective Lagrangian. For the complete matching at the scale
$\mu_0$, at the leading order in $1/M_W$ and including
$\mathcal{O}(\alpha_e)$ terms, one has to retain all the
operators from Eq.~(\ref{4operators}), and supplement
them with $Q_2^\dagger$ defined in Eq.~(\ref{q2def}).
The cancellation of spurious infrared divergence effects requires also the
inclusion of additional evanescent operators (see Appendix A
of Ref.~\cite{Bobeth:2013tba}). The renormalization constants of all
resulting Wilson coefficients are then calculated in a way analogous
to the $C_A^{(m,0)}$ calculation.
\begin{figure}[t]
\begin{center}
\includegraphics[width=15.5 cm]{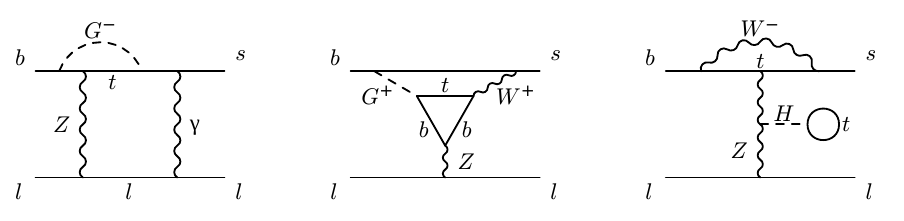}
\caption{Examples of diagrams contributing to the
$\mathcal{O}(\alpha_e)$ EW correction on the SM side of the matching
equation~(\ref{matching eq}). Diagrams from Fig.~1 of
Ref.~\cite{Bobeth:2013tba}.\label{fig4}}
\end{center}
\end{figure}

Examples of diagrams contributing to $C_A^{(0,1)}(\mu_0)$ on
the SM side are shown in Fig.~\ref{fig4}. Before this correction can
be extracted, the UV divergences on the SM side have to be
renormalized. For a discussion of different renormalization
schemes for the electroweak boson and top quark
masses, we refer to the original article
\cite{Bobeth:2013tba}. Here, we will continue the analysis in the OS-2
scheme defined therein, as it was used in the subsequent
phenomenological analysis~\cite{Bobeth:2013uxa}. In this scheme, all
the QCD corrections to mass renormalization constants are defined in
the $\overline{\text{MS}}$, but the $\mathcal{O}(\alpha_e)$
corrections to $Z_{m_t}$, $Z_{M_W}$ and $Z_{M_Z}$ are defined on shell. In
practice, the calculation was first done fully in the
$\overline{\text{MS}}$ scheme, and the finite terms in these three
renormalization constants were subsequently added to the renormalized
results.

The value of $C_A$ extracted at the scale $\mu_0$ must then be
RG-evolved down to the $\mu_b$ scale of the $B_s\to\mu^+\mu^-$
decay. The renormalized Wilson coefficients of the extended set of
effective operators are related to the bare ones through a relation
similar to Eq.~(\ref{bareC}):
\begin{equation}
C_j^{(b)}Q_j^{(b)}=Z_qZ_l\sum\limits_kC_kZ_{kj}Q_j\Longrightarrow C_j^{(b)}=\sum\limits_kC_kZ_{kj}.
\end{equation}
The one-loop RGE for Wilson coefficients $\vec{C}$ has a general form
\begin{equation}
\label{RGE}
\mu\frac{d}{d\mu}\vec{C} = \left( \tilde{\alpha}_s\gamma_s^{(0)} + \tilde{\alpha}_e\gamma_e^{(0)} \right)^T  \vec{C},
\end{equation}
where the Anomalous Dimension Matrices (ADMs)
$\gamma_s^{(0)}$ and $\gamma_e^{(0)}$ are obtained from the
renormalization constants of Wilson coefficients. Once we restrict to
non-evanescent operators only, the necessary
$\text{MS}$-scheme relation reads
\begin{equation}
Z_{kj}=\delta_{kj}+ \frac{1}{2\epsilon}
\left[ \tilde{\alpha}_s \gamma_s^{(0)} + \tilde{\alpha}_e \gamma_e^{(0)} \right]_{kj} +\mathcal{O}\left(\tilde{\alpha}_s^2,\tilde{\alpha}_e^2,\tilde{\alpha}_s\tilde{\alpha}_e\right).
\end{equation} 
For the RGE~(\ref{RGE}) to close, one has to extend the set of
operators in the effective Lagrangian further (see the discussion
under Eq.~(15) in Ref.~\cite{Bobeth:2013tba}).
Analytical expressions for the evolution operator associated
with Eq.~(\ref{RGE}) can be found, e.g., in
Ref.~\cite{Huber:2005ig}. The details and results of the
numerical solution can be found in Appendix B of
Ref.~\cite{Bobeth:2013tba}.

\subsection{Power-enhanced QED corrections}\label{subsection:EW2}

\begin{figure}[t]
\begin{center}
\includegraphics[width=15.5 cm]{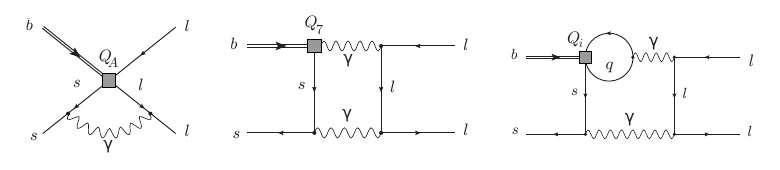}
\caption{Examples of diagrams with virtual photon exchanges
that are responsible for the power-enhanced QED correction in
Eq.~(\ref{eta}). Diagrams from Fig.~1 of
Ref.~\cite{Beneke:2017vpq}.\label{fig5}}
\end{center}
\end{figure}

Within the effective theory framework, the $S$-matrix element
corresponding to the $B_s\to\mu^+\mu^-$ decay receives
$\mathcal{O}(\alpha_e)$ contributions from diagrams with
tree-level Wilson coefficients and a virtual photon exchanged between
the fermions. In particular, there is a class of diagrams with a photon
emitted from the spectator $s$ quark and absorbed by one of the
outgoing muons (see Fig.~\ref{fig5}). The leading terms of the
power series in $\{m_s,m_\mu,\Lambda_{\text{QCD}}\}/m_b$ of this
correction were computed in Ref.~\cite{Beneke:2017vpq}. It was
observed that its helicity suppression by the factor of $r^2$ in the
tree-level branching ratio (\ref{simplifiedbr}) is partially lifted
due to a relative enhancement by $M_{B_s}/\Lambda_{\text{QCD}}$.

Effects of hard virtual photons with energies and virtualities
above the $\mu_b$ scale are contained in the Wilson coefficients
$C_i$. The remaining virtual photons need to be taken into account
in the physical matrix elements that are evaluated at the scale
$\mu_b$. In particular, virtual photons exchanged in the
diagrams in Fig.~\ref{fig5} probe the inner structure of the $B_s$
meson, smearing the annihilation point of the valence quarks over a
distance $\mathcal{O}\left((M_{B_s}\Lambda_{\text{QCD}})^{-\frac{1}{2}}\right)$, which
corresponds to inverse virtuality of the off-shell $s$
quark. Such long-distance QCD effects cannot be parameterized
solely by $f_{B_s}$, as at the leading order. Instead, one has to
estimate effects of matrix elements like
\begin{equation}
\braket{0|\!\int\! d^dx\;\mathbf{T}[[\bar{\mu}\gamma^\alpha\mu](x)Q_A(0)]|B_s},
\end{equation}
where \textbf{T} is the time-ordering operator. They involve the $B$
meson light-cone distribution amplitude~\cite{Beneke:1999br} and its
first logarithmic moments.

Virtual photon exchanges leading to power-enhanced QED
corrections were thoroughly studied in the formalism of the
Soft-Collinear Effective Theory by Beneke, Bobeth and Szafron
in Ref.~\cite{Beneke:2019slt}. Their effect can be included in
the SM prediction through a replacement
\begin{equation}
\overline{\mathcal B}_{s\mu} \rightarrow\eta_{\scriptscriptstyle BBS} \overline{\mathcal B}_{s\mu}~~,
\end{equation}
with 
\begin{equation} \label{eta}
\eta_{\scriptscriptstyle BBS}=0.995^{+0.003}_{-0.005}.
\end{equation}
The main uncertainty in $\eta_{\scriptscriptstyle BBS}$ comes
from poorly known values of hadronic
parameters~\cite{Beneke:2011nf}. We have extracted the numerical value
of $\eta_{\scriptscriptstyle BBS}$ as well as its uncertainty in
Eq.~(\ref{eta}) from Eqs.~(8.8) and (8.10) of
Ref.~\cite{Beneke:2019slt}. One can observe there that the relatively
small central value of the correction ($-0.5\%$) arises as an effect
of partial cancellation between potentially unrelated contributions
from the $Q_V$ and $Q_7^\dagger$ operators. Thanks to this
cancellation, the overall effect remains below the $\pm 1.5\%$
non-parametric uncertainty estimated in Ref.~\cite{Bobeth:2013uxa}.
That uncertainty was primarily due to unknown $\mathcal{O}(\alpha_e)$
effects, although the possibility of their power enhancement remained
unknown at the time.

The contribution of $Q_7^\dagger$ to the considered corrections
was reanalyzed in Ref.~\cite{Feldmann:2022ixt}. Despite several
differences in the analytical expressions with respect to the earlier
analysis of Ref.~\cite{Beneke:2019slt}, the numerical results of the
two papers remain in qualitative agreement, and no modification of the
factor $\eta_{\scriptscriptstyle BBS}$ in Eq.~(\ref{eta}) is
necessary.

\section{Numerical analysis}\label{section:num}

In this section, we update the SM prediction for
$\overline{\mathcal B}_{s\mu}$ based on
Eq.~(\ref{simplifiedbr}), including the complete $\mathcal{O}(\alpha_s^2)$ and
$\mathcal{O}(\alpha_e)$ corrections to $C_A$, as well as the QED correction
factor in Eq.~(\ref{eta}). In practice, it is sufficient to use the
semi-numerical expressions from Eq.~(6) of Ref.~\cite{Bobeth:2013uxa},
which gives
\begin{equation} \label{bsmu}
\overline{\mathcal B}_{s\mu}^{SM} \times 10^9 = (3.65 \pm 0.06) 
\left( \frac{M_t[{\rm GeV}]}{173.1} \right)^{\!\! 3.06}
\left( \frac{\alpha_s(M_Z)}{0.1184} \right)^{\!\!\! -0.18} R_s\, \eta_{\scriptscriptstyle BBS} \,,
\end{equation}
where
\begin{equation} \label{rs}
R_s = \left( \frac{f_{B_s}[{\rm MeV}]}{227.7} \right)^{\! 2}\!
\left( \frac{|V_{cb}|}{0.0424} \right)^{\! 2}\!
\left( \frac{|V_{tb}^\star V_{ts}/V_{cb}|}{0.980} \right)^{\! 2}
\frac{\tau_H^s\,[{\rm ps}]}{1.615}\,.
\end{equation}
In the above expressions, all the explicitly displayed input
parameters are normalized to their 2013 central values. In
Table~\ref{tab1}, we update these central values together with the
corresponding uncertainties. All the remaining parameters are retained
unaltered with respect to Table~I of Ref.~\cite{Bobeth:2013uxa}, as
their update would not affect $\overline{\mathcal B}_{s\mu}$ in a
noticeable manner -- they are either very precisely measured or have
little effect on $\overline{\mathcal B}_{s\mu}$.
\begin{table}[t] 
\begin{center}
\caption{Numerical values of the updated input parameters.\label{tab1}}
\begin{tabular}{ccccccc}\hline
\textbf{Parameter}           && \textbf{Value}              && \textbf{Unit}       && \textbf{Ref.}\\\hline
$f_{B_s}$                    && 230.3\,(1.3)         && MeV          && \cite{FlavourLatticeAveragingGroupFLAG:2021npn,Bazavov:2017lyh,ETM:2016nbo,Dowdall:2013tga,Hughes:2017spc}\\
$|V_{cb}|\times 10^3$        && 41.97\,(48)          && -            && \cite{Finauri:2023kte}\\
$|V_{tb}^*V_{ts}/V_{cb}|$    && 0.9820\,(4)          && -            && derived from Ref.~\cite{Charles:2004jd}\\
$\tau_H^s$                   && 1.622\,(8)    && ps           && \cite{HFLAV:2022esi}\\
$M_t$                        && 172.57\,(29)  && GeV          && \cite{PDG2024}\\
$\alpha_s(M_Z)$              && 0.1180\,(9)          && -            && \cite{PDG2024}\\\hline
\end{tabular}
\end{center}
\end{table}
  
As already mentioned in Eq.~(\ref{brsm}), we find
$\overline{\mathcal B}_{s\mu}^{SM} = (3.64 \pm 0.12) \times 10^{-9}$.  The
overall uncertainty is now almost a factor of two smaller than found
in Ref.~\cite{Bobeth:2013uxa}, while the central value remains almost
unchanged. The latter fact can be attributed to an approximate
cancellation of shifts stemming from the parameter updates and
$\eta_{\scriptscriptstyle BBS}$, as in the following sum:
\begin{equation}
+ 2.3\% (f_{B_s})
- 1.6\% (CKM)
+ 0.4\% (\tau_H^s)
- 0.9\% (M_t)
+ 0.1\% (\alpha_s)
- 0.5\% (\eta_{\scriptscriptstyle BBS})
\simeq -0.3\%.
\end{equation}
As far as the uncertainty breakdown is concerned, its current version
is compared to the 2013 one~\cite{Bobeth:2013uxa} in Table~\ref{tab2}.
In its last column, the uncertainties are combined in quadrature.
\begin{table}[t] 
\caption{The current uncertainty breakdown in $\overline{\mathcal B}_{s\mu}^{SM}$, as compared to the 2013 one. \label{tab2}}
\begin{tabular}{lccccccccc}\hline
& $f_{B_s}$ 
& CKM 
& $\tau_H^s$ 
& $M_t$ 
& $\alpha_s$
& $\eta_{\scriptscriptstyle BBS}$
& other      
& non-        
& $\sum$
\\[-1mm]
&&&&&&&& parametric &\\\hline
2024 [this paper]
& $1.1$\%            
& $2.3\%$            
& $0.5$\%            
& $0.5$\%            
& $0.1$\%            
& $0.5$\%            
& $< 0.1$\%          
& $1.5$\%            
& $3.2$\%            
\\
2013~\cite{Bobeth:2013uxa} 
& $4.0$\%
& $4.3\%$ 
& $1.3$\%
& $1.6$\%
& $0.1$\%
& $0.0$\%
& $< 0.1$\%
& $1.5$\%
& $6.4$\%
\\\hline
\end{tabular}
\end{table}
One can observe a significant improvement in the first four columns
where the dominant parametric uncertainties originate from, in
particular in the case of $f_{B_s}$ that is determined on the
lattice. As far as the top-quark mass $M_t$ is concerned, let us
recall that the PDG~\cite{PDG2024} value is treated as the on-shell
mass, which is not strictly correct. However, the overall
non-parametric uncertainty of $\pm 1.5\%$ is understood to contain a
contribution from such an approximation. As it is evident from
Tables~\ref{tab1} and \ref{tab2}, a $300\;{\rm MeV}$ shift in $M_t$
implies a $0.5\%$ shift in $\overline{\mathcal B}_{s\mu}^{SM}$.

At present, the most important uncertainty originates from $|V_{cb}|$,
in which case we use the inclusive determination
only~\cite{Finauri:2023kte}. A combination with exclusive
determinations would not lead to an improvement, given the persistent
discrepancy between the inclusive and exclusive results~\cite{PDG2024}. Our
preference is the same as in Ref.~\cite{Bobeth:2013uxa}, i.e.\ we
treat the inclusive determination as theoretically cleaner, and more
reliable.

As already mentioned in Ref.~\cite{Bobeth:2013uxa}, one can get rid of
$|V_{cb}|$ in the ratio of $\overline{\mathcal B}_{s\mu}$ to the
measured $B_s^{(H)}$-$B_s^{(L)}$ mass difference, at the cost of
introducing an extra uncertainty from lattice determinations of the
``bag parameter'' $B_{B_s}$. Such an approach is likely to become
relevant once the experimental accuracy in $\overline{\mathcal
B}_{s\mu}$ (currently $8.1\%$ in Eq.~(\ref{experiment})) becomes
closer to the theoretical one.

\section{Conclusions}\label{section:conc}

In this review, we analyzed the current SM prediction for the
branching ratio of the rare $B_s\to\mu^+\mu^-$ decay. This channel
continues to be among the most promising candidates for detecting BSM
physics without direct production of new particles, due to its SM
suppression and possible BSM enhancements.

The SM analysis is, to a very good approximation, contained in the
perturbative calculation of the Wilson coefficient
$C_A$, and the lattice calculation of the long-distance QCD
parameter $f_{B_s}$. The perturbative part was already complete up to
and including next-to-next-to-leading QCD and
next-to-leading EW effects.

The currently dominant theoretical uncertainty originates from the
CKM-matrix element $|V_{cb}|$. The next-to-dominant uncertainty is
already non-parametric, stemming mainly from the unknown higher-order
electromagnetic corrections at the scale $\mu_b$. They depend on
non-perturbative effects that are not contained in $f_{Bs}$. The
current result for $\overline{\mathcal B}_{s\mu}^{SM}$ changes by
around $0.3\%$ when $\mu_b$ is varied between $m_b/2$ and $2m_b$.
However, since it provides a lower bound only on the possible size of
unknown electromagnetic effects, the actual uncertainty estimate
should be somewhat more conservative. Here, we have retained the
overall non-parametric uncertainty at the same level as in 
Ref.~\cite{Bobeth:2013uxa}, namely $\pm 1.5\%$. 

The experimental error is currently much larger, around 8\%, which
sets a limit on the power of $B_s\to\mu^+\mu^-$ as a means for testing
various BSM theories. The situation is expected to improve with time,
when higher statistics get collected at the LHC and future
experiments.

A final message that we would like to share with the reader is as
follows. Our numerical result in Eq.~(\ref{brsm}) will become outdated
as soon as any of the input parameters gets determined in a new
analysis. Performing another update of $\overline{\mathcal
B}_{s\mu}^{SM}$ does not require being an expert. It is sufficient to
substitute new inputs into the simple formula~(\ref{bsmu}).

\ \\
{\bf Acknowledgements:}
This work was supported by the National Science Center,
Poland, under the research project 2020/37/B/ST2/02746.

\def\thesection{Appendix A}
\def\theequation{A\arabic{equation}}
\setcounter{equation}{0}
\setcounter{table}{0}
\section{The branching ratio formula}\label{section:Appendix A}

In this appendix, we sketch the derivation of the branching ratio
formula~(\ref{BR2HDM}) that holds in models with SM-like ${\rm CP}$-violation,
including the SM and the 2HDM. We work at the leading order in QED
throughout, i.e. the final-state muons are understood to be emitted
directly from the operator vertex ($Q_A$, $Q_P$ or $Q_S$).

Let $\ket{B_s}$ and $\ket{\overline{B}_s}$ denote the meson flavour
eigenstates with valence quarks $\bar{b}s$ and $b\bar{s}$,
respectively. We fix conventions in their overall phases by demanding
that~
${\rm CP} \ket{B_s} = \ket{\overline{B}_s}$,~ and 
${\rm CPT} \ket{B_s} = \ket{\overline{B}_s}$. Once this is done,
the heavier ($H$) and lighter ($L$) mass eigenstates (see section 13
of Ref.~\cite{PDG2024}) can respectively be written as
\begin{equation} \label{HLbasis}
\ket{B_s^{(H)}} = \frac{1}{\sqrt{2}|N|} \left( N^* \ket{B_s} - N \ket{\overline{B}_s} \right),
\hspace{1cm}
\ket{B_s^{(L)}} = \frac{1}{\sqrt{2}|N|} \left( N^* \ket{B_s} + N \ket{\overline{B}_s} \right),
\end{equation}
where $N$ has been defined in Eq.~(\ref{N def}). In the limit of no
${\rm CP}$-violation ($N=|N|$), $B_s^{(H)}$ and $B_s^{(L)}$ are
${\rm CP}$-odd and ${\rm CP}$-even, respectively.

From the form of the lepton currents in Eq.~(\ref{4operators}),
we observe that the $Q_A$ and $Q_P$ interactions can
lead to production of ${\rm CP}$-odd lepton pairs only (in
the CM frame), while $Q_S$ can lead to production of
${\rm CP}$-even pairs only. In the $Q_A$ case, it follows from the
fact that the timelike component of the lepton current is ${\rm
CP}$-odd (see, e.g., the table below Eq.~(3.150) of
Ref.~\cite{Peskin:1995ev}), while the spacelike components play no
role, as they get contracted with vanishing spacelike components of
the meson momentum (see Eq.~(\ref{fdef})).

Let us now show that even in the presence of SM-like ${\rm
CP}$-violation, ($N \neq |N|$ but real Wilson coefficients), the
operators $Q_A$ and $Q_P$ have no effect on dimuonic decays of
$B_s^{(L)}$, while $Q_S$ has no effect on such decays of
$B_s^{(H)}$. When the ``${\mathrm h.c.}$'' terms in the Lagrangian
(\ref{main lag}) are taken into account, the leading $Q_A$
contribution to the $B_s^{(L)}$ decay amplitude is proportional to
\begin{equation} \label{ALzero}
\bra{\mu^+\mu^-} N Q_A + N^* Q_A^\dagger \ket{B_s^{(L)}} =
\frac{|N|}{\sqrt{2}} \left( \bra{\mu^+\mu^-} Q_A \ket{B_s} + \bra{\mu^+\mu^-} Q_A^\dagger \ket{\overline{B}_s} \right),
\end{equation}
where the matrix elements on the r.h.s.\ are the only two that do not
vanish due to flavour conservation in QCD. Next, we observe that
\begin{equation} 
\bra{\mu^+\mu^-} Q_A^\dagger \ket{\overline{B}_s} = \bra{\mu^+\mu^-} ({\rm CP})^\dagger Q_A {\rm CP} \ket{\overline{B}_s} = -\bra{\mu^+\mu^-} Q_A \ket{B_s}, 
\end{equation}
where we have taken advantage of the fact that the dimuon state is
${\rm CP}$-odd, as argued above. Consequently, the sum of the two
matrix elements on the r.h.s.\ of Eq.~(\ref{ALzero}) vanishes. An
identical reasoning holds for $Q_P$. As far as $Q_S$ and $B_s^{(H)}$
are concerned, we proceed by analogy:
\begin{equation} \label{SHzero}
\bra{\mu^+\mu^-} N Q_S + N^* Q_S^\dagger \ket{B_s^{(H)}} =
\frac{|N|}{\sqrt{2}} \left( \bra{\mu^+\mu^-} Q_S \ket{B_s} - \bra{\mu^+\mu^-} Q_S^\dagger \ket{\overline{B}_s} \right),
\end{equation}
\begin{equation} 
\bra{\mu^+\mu^-} Q_S^\dagger \ket{\overline{B}_s} = \bra{\mu^+\mu^-} ({\rm CP})^\dagger Q_S {\rm CP} \ket{\overline{B}_s} = \bra{\mu^+\mu^-} Q_S \ket{B_s}, 
\end{equation}
where this time the dimuon state is ${\rm CP}$-even.  Consequently,
the difference of the two matrix elements on the r.h.s.\ of
Eq.~(\ref{SHzero}) vanishes. As a by-product of the above reasoning,
we can simplify the non-vanishing matrix elements as follows:
\begin{equation} \label{melems}
\begin{split}
\bra{\mu^+\mu^-} N Q_{A,P} + N^* Q_{A,P}^\dagger \ket{B_s^{(H)}} &= |N| \sqrt{2} \bra{\mu^+\mu^-} Q_{A,P} \ket{B_s},\\
\bra{\mu^+\mu^-} N Q_S + N^* Q_S^\dagger \ket{B_s^{(L)}} &= |N| \sqrt{2} \bra{\mu^+\mu^-} Q_S \ket{B_s}.
\end{split}
\end{equation}

Both at the LHC and at $e^+ e^-$ machines, the production rates of
$B_s$ and $\overline{B}_s$ are practically equal. Thus, to a very good
approximation, the heavy and light mass eigenstates are produced in
the same quantities, as can be seen by inverting the relation
($\ref{HLbasis}$). Since the decay products of $B_s^{(H)}$ (${\rm
CP}$-odd dimuons) and $B_s^{(L)}$ (${\rm CP}$-even dimuons) do not
interfere, the average time-integrated branching ratio is simply given by
\begin{equation} \label{avbr}
\overline{\mathcal B}_{s\mu} =\frac{1}{2}\left(\frac{\Gamma[B^{(H)}_s\rightarrow\mu^+\mu^-]}{\Gamma_H^s}+\frac{\Gamma[B^{(L)}_s\rightarrow\mu^+\mu^-]}{\Gamma_L^s}\right).
\end{equation}
In each of the two cases, the decay rate is given by the well-known formula
\begin{equation} \label{phase space}
\Gamma[B^{(H,L)}_s\rightarrow\mu^+\mu^-] =
\frac{1}{2 M_{B_s}} \int dPS_2 \left| \overline{{\mathcal M}}^{(H,L)} \right|^2
= \frac{\beta}{16\pi M_{B_s}} \left| \overline{{\mathcal M}}^{(H,L)} \right|^2,
\end{equation} 
with
\begin{equation} 
dPS_2 = \frac{1}{4\pi^2}\, d^4 k_+ \, d^4 k_- \,
\delta(k_+^2-m_\mu^2) \theta(k^0_+)\, \delta(k_-^2-m_\mu^2) \theta(k^0_-)\, \delta^{(4)}(p-k_+-k_-).
\end{equation} 
Here, ${\mathcal M}^{(H,L)}$ are the corresponding invariant matrix
elements, $\beta$ has been defined below Eq.~(\ref{simplifiedbr}), and
we have neglected the tiny mass splitting between $B_s^{(H)}$ and
$B_s^{(L)}$. Summing over spins of the final-state muons is understood in
Eq.~(\ref{phase space}). The two-body phase-space integral is trivial,\footnote{
see section 3.2 of Ref.~\cite{Lang:2023bdy}}
as $\left| \overline{{\mathcal M}}^{(H,L)} \right|^2$ is constant in
the integration domain due to rotational symmetry in the decaying
scalar rest frame.

The $Q_A$ (and $Q_A^\dagger$) contribution to ${\mathcal M}^{(H)}$ reads
\begin{eqnarray} 
{\mathcal M}^{(H)}_A &=& i C_A |N| \sqrt{2}\, e^{ipx} \bra{0} j^\alpha_A(x) \ket{B_s(p)}\, \bar{u}(k_-) \gamma^\alpha \gamma_5\, v(k_+)\nonumber\\[2mm]
&=& -f_{B_s} C_A |N| \sqrt{2}\, \bar{u}(k_-) \slashed{p}  \gamma_5\, v(k_+), \label{HAmelem}
\end{eqnarray}
where the identity (\ref{melems}) has already been taken into
account. Moreover, the $\ket{B_s}$ state normalization has been
adjusted to the one that is conventionally used in the decay-constant
definition (\ref{fdef}), namely
\begin{equation} \label{normB}
\braket{B_s(q)|B_s(p)} = 2 p_0 (2\pi)^3 \delta^{(3)}\left(\vec{p}-\vec{q}\right).
\end{equation}

To verify that the global normalization in Eq.~(\ref{HAmelem}) is
correct, one can begin with the relevant $S$ matrix element
\begin{equation}
\braket{S}\equiv{}_{\rm out}\!\braket{\mu^+\mu^-|B_s^{(H,L)}}_{\rm in}=\braket{l^+l^-|\mathbf{T}[\exp(i\int d^4x\, \mathcal{L}_{\rm int})]|B_s^{(H,L)}}.
\end{equation}
In its evaluation, the $B_s$ meson hadronic structure cannot be
treated in a perturbative manner. Therefore, in the definition of the
in/out states in the above equation, the interaction part of the
Lagrangian $\mathcal{L}_{\rm int}$ that is switched off at timelike
infinities, consists only of the weak part $N\sum_nC_nQ_n+{\mathrm
h.c.}$. It is assumed that non-perturbative QCD effects have been
solved beforehand, and are contained in $\ket{B_s^{(H,L)}}_{\rm in}$ that
is treated as an asymptotic state from the weak interaction
perspective.

In practice, the right normalization in Eq.~(\ref{HAmelem}) can be
determined with possibly least effort by considering an analogy with a
certain purely perturbative theory. Suppose the muons interact with
a massive real pseudoscalar $\phi$ via a dimension-five operator
$\widetilde{{\mathcal L}}_{\rm int} = \frac{\lambda}{M}
(\partial^\alpha \phi) \bar\mu \gamma_\alpha \gamma_5
\mu$.  Once the tree-level invariant matrix element ${\mathcal
M}$ for the decay $\phi \to \mu^+ \mu^-$ is quickly determined, it
should be expressed in terms of the free-field matrix element~
$e^{ipx} \bra{0} \frac{\lambda}{M} \partial^\alpha \phi(x) \ket{\phi(p)}$,
with the free-particle state $\ket{\phi(p)}$ normalized as in
Eq.~(\ref{normB}). Finally, replacing
\begin{displaymath}
e^{ipx} \bra{0} \frac{\lambda}{M} \partial^\alpha \phi(x) \ket{\phi(p)} 
\hspace{1cm} \mbox{by} \hspace{1cm}
C_A |N| \sqrt{2}\, e^{ipx} \bra{0} j^\alpha_A(x) \ket{B_s(p)}, 
\end{displaymath}
(see Eq.~(\ref{melems})), one obtains Eq.~(\ref{HAmelem}) with the proper
overall normalization.

The r.h.s.\ of Eq.~(\ref{HAmelem}) can be further simplified by noticing that
\begin{equation}
\bar{u} \slashed{p}  \gamma_5\, v =
\bar{u} (\slashed{k}_- + \slashed{k}_+) \gamma_5\, v =
\bar{u} (\slashed{k}_- \gamma_5 - \gamma_5 \slashed{k}_+) v = 2 m_\mu \bar{u} \gamma_5 v,
\end{equation}
where the identities
$\bar{u}(k_-) \slashed{k}_- = m_\mu \bar{u}(k_-)$,~
and~ $\slashed{k}_+ v(k_+) = -m_\mu v(k_+)$~
have been used in the last step. Let us note that our expression would
vanish in the absence of $\gamma_5$, which is related to the vanishing
divergence of the vectorlike current, the same one that enters the QED
interactions of muons and photons. This is precisely the reason why we
were allowed to skip $Q_V$ in our initial operator list in Eq.~(\ref{4operators}).

Let us now turn to the operators $Q_P$ and $Q_S$. The pseudoscalar
current $j_P = \bar{b}\gamma_5 s$ that is present in both of them can
be eliminated in favour of the axial current $j_A^\mu $ using
Equations of Motion (EoM) for the $s$- and $\bar{b}$-quark operator
fields:
\begin{equation} \label{eom}
\begin{split}
[i \slashed{\partial} - g\slashed{A}^a T^a - e\slashed{A} - m_s]s &\underset{\text{EoM}}{=}\mathcal{O}(M_W^{-2}),\\
\bar{b}[i\overset{\leftarrow}{\slashed{\partial}} + g\slashed{A}^a T^a + e\slashed{A}+m_b] & \underset{\text{EoM}}{=}\mathcal{O}(M_W^{-2}),
\end{split}
\end{equation}
where $g$ and $e$ are the QCD and QED couplings respectively,
$A_\mu^a$ is the gluon field, $A_\mu$ is the photon field, and $T^a$
are the SU(3) generators. The $\mathcal{O}(M_W^{-2})$ effects on the
r.h.s.\ of the above equations stand for higher-order weak-interaction
effects that stem from the operators $Q_n$ in Eq.~(\ref{main
lag}). Such effects will be neglected below, as we are going
to use the EoM to transform the weak operators $Q_P$ and $Q_S$, while
we work at the leading order in weak interactions. Operators that
vanish by the EoM will commonly be denoted by $\boxed{E}$. They
can be skipped in evaluation of observables at the leading order in
weak interactions, as their physical matrix elements vanish.

To proceed, we multiply the first equation in (\ref{eom}) by
$(\bar{b}\gamma_5)$ from the left, the second one by $(\gamma_5s)$
from the right, take their difference, rearrange, and obtain
\begin{equation}
\partial_\alpha j_A^\alpha = i(m_b+m_s)j_P + \boxed{E}.
\end{equation}
Using the above identity, one can express the following total derivative as
\begin{equation}
\partial_\alpha[(\bar{l}l)j_A^\alpha]=j_A^\alpha \partial_\alpha[\bar{l}l]+i(m_b+m_s)Q_S+\boxed{E}.
\end{equation}
Consequently,
\begin{equation} \label{replQS}
Q_S = \frac{(i\partial_\alpha[\bar{l}l])j_A^\alpha}{m_b+m_s} + \boxed{E} + \boxed{T},
\end{equation}
where $\boxed{T}$ commonly denotes total derivatives of operators
that are invariant under QCD and QED gauge transformations. Similarly,
\begin{equation} \label{replQP}
Q_P = \frac{(i\partial_\alpha[\bar{l}\gamma_5l])j_A^\alpha}{m_b+m_s} + \boxed{E} + \boxed{T}.
\end{equation}
Since $\boxed{T}$ in the Lagrangian has no effect on physical
observables, we are allowed to replace $Q_S$ and $Q_P$ by the first
(explicit) terms in the above equations. After such replacements, the
same quark current shows up in $Q_A$, $Q_S$ and $Q_P$, which means
that only a single non-perturbative quantity, namely $f_{B_s}$, is
sufficient to describe their physical matrix elements at the leading
order in QED.

Using Eqs.~(\ref{replQS})--(\ref{replQP}), and performing calculations
along the same way as in the case of $Q_A$, one obtains
\begin{equation}
\mathcal{M}_P^{(H)} = \frac{M^2_{B_s} f_{B_s}}{m_b+m_s} C_P |N| \sqrt{2} \, \bar{u}(k_-)\gamma_5\, v(k_+),
\end{equation}
and
\begin{equation}
\mathcal{M}_S^{(L)} = \frac{M^2_{B_s} f_{B_s}}{m_b+m_s} C_S |N| \sqrt{2} \, \bar{u}(k_-) v(k_+).
\end{equation}
The total invariant matrix element for $B_s^{(H)}$ reads
\begin{equation}
\mathcal{M}^{(H)} = \mathcal{M}_P^{(H)} + \mathcal{M}_A^{(H)} = 
\left( \frac{M_{B_s} C_P}{m_b+m_s} - \frac{2 m_\mu C_A}{M_{B_s}} \right)
M_{B_s} f_{B_s} |N| \sqrt{2} \, \bar{u}(k_-)\gamma_5\, v(k_+).
\end{equation}

It remains to take the moduli squared and perform the sums over spins:
\begin{equation}
\begin{split}
\sum_{\rm spins} |\bar{u} \gamma_5\, v|^2 &= -{\rm Tr}\left[ (\slashed{k}_- + m_\mu) \gamma_5  (\slashed{k}_+ - m_\mu) \gamma_5 \right] = 2 M_{B_s}^2,\\
\sum_{\rm spins} |\bar{u} v|^2 &= {\rm Tr}\left[ (\slashed{k}_- + m_\mu) (\slashed{k}_+ - m_\mu) \right] = 2 M_{B_s}^2 \beta^2.
\end{split}
\end{equation}
Finally, after substitution to Eq.~(\ref{phase space}), and then (\ref{avbr}), we end up with the branching ratio formula quoted in Eq.~(\ref{BR2HDM}).

In BSM models with generic (not SM-like) ${\rm CP}$-violation, the
Wilson coefficients $C_A$, $C_P$ and $C_S$ are not necessarily
real. Moreover, the mass eigenstates are not necessarily given by
Eq.~(\ref{HLbasis}), as the phase factors may differ from $N/|N|$ and
$N^*/|N|$.  The latter effect can be described by introducing an extra
complex phase $\phi_s^{BSM} \simeq \phi_s^{c\bar c s} - \arg[(V^*_{ts}
V_{tb})^2]$, see section 2.2 of Ref.~\cite{Buras:2013uqa}. In such
models, the branching ratio formula generalizes to
\begin{equation}
\label{brFormula}
\overline{\mathcal B}_{s\mu} =\frac{|N|^2M_{B_s}^3f_{B_s}^2}{8\pi\Gamma_H^s}\beta\left[|rC_A-uC_P|^2F_P+|u\beta C_S|^2F_S\right]+\mathcal{O}(\alpha_{e}),
\end{equation}
with
\begin{equation}
\begin{split}
F_P&\equiv1-\frac{\Gamma_L^s-\Gamma_H^s}{\Gamma_L^s}\sin^2\left[\frac{1}{2}\phi_s^{\rm BSM}+\arg(rC_A-uC_P)\right],\\
F_S&\equiv1-\frac{\Gamma_L^s-\Gamma_H^s}{\Gamma_L^s}\cos^2\left[\frac{1}{2}\phi_s^{\rm BSM}+\arg(rC_S)\right].
\end{split}
\end{equation}
The above expressions for $F_P$ and $F_S$ have been derived from
the results of Refs.~\cite{DeBruyn:2012wk,DeBruyn:2012wj} where an
interesting discussion concerning time-dependent observables can be
found.

\def\thesection{Appendix B}
\def\theequation{B\arabic{equation}}
\def\thetable{B\arabic{table}}
\setcounter{equation}{0}
\setcounter{table}{0}
\section{Numerical update for $\overline{\mathcal B}_{d\mu}^{SM}$}\label{section:Appendix B}
\begin{table}[t]
\begin{center}
\caption{Numerical values of the extra input parameters that matter for $\overline{\mathcal B}_{d\mu}$. \label{tab3}}
\begin{tabular}{ccccccc}\hline
\textbf{Parameter}           && \textbf{Value}       && \textbf{Unit} && \textbf{Ref.}\\\hline
$f_{B_d}$                    && 190.0\,(1.3)         && MeV           && \cite{FlavourLatticeAveragingGroupFLAG:2021npn,Bazavov:2017lyh,ETM:2016nbo,Dowdall:2013tga,Hughes:2017spc}\\
$|V_{tb}^*V_{td}|$           && 0.00851\,(10)        && -             && derived from Ref.~\cite{Charles:2004jd}\\
$\tau_{\rm av}^d$            && 1.517\,(4)           && ps            && \cite{HFLAV:2022esi}\\\hline
\end{tabular}
\end{center}
\end{table}

Here, we present the current SM prediction for $\overline{\mathcal
B}_{d\mu}$, i.e.\ the average time-integrated branching ratio of $B^0
\to \mu^+\mu^-$. The necessary formula is analogous to
Eq.~(\ref{bsmu}), obtained by combining the semi-numerical expression
from Eq.~(7) of Ref.~\cite{Bobeth:2013uxa} with
$\eta_{\scriptscriptstyle BBS}$ (\ref{eta}) derived from
Ref.~\cite{Beneke:2019slt}. It reads
\begin{equation} \label{bdmu}
\overline{\mathcal B}^{\rm SM}_{d\mu} \times 10^{10} = (1.06 \pm 0.02) 
\left( \frac{M_t[{\rm GeV}]}{173.1} \right)^{\!\! 3.06}
\left( \frac{\alpha_s(M_Z)}{0.1184} \right)^{\!\!\! -0.18} R_d\, \eta_{\scriptscriptstyle BBS} \,,
\end{equation}
where
\begin{equation} \label{rd}
R_d = \left( \frac{f_{B_d}[{\rm MeV}]}{190.5} \right)^{\! 2}\!
\left( \frac{|V_{tb}^\star V_{td}|}{0.0088} \right)^{\! 2}
\frac{\tau_{\rm av}^d\,[{\rm ps}]}{1.519}\,.
\end{equation}
As input, we need three more parameters in addition to those already
listed in Table~\ref{tab1}, namely the decay constant $f_{B_d}$ of the
$B^0$ meson, the average lifetime of this meson $\tau_{\rm av}^d$, and
the relevant CKM factor $|V_{tb}^*V_{td}|$. Their current values are
listed in Table~\ref{tab3}.  Our use of the well-measured $\tau_{\rm
av}^d$ instead of $\tau_H^d$ is a good approximation thanks to the
very small width difference predicted in the SM for the
$B^0$-${\overline B}^0$ system: $\Delta \Gamma_d/(2 \Gamma^d_{\rm av})
= 0.00172(46)$~\cite{Asatrian:2020zxa}.

After substituting all the numerical inputs to Eq.~(\ref{bdmu}), we
find
\begin{equation} 
\overline{\mathcal B}^{\rm SM}_{d\mu} = (9.71 \pm 0.33) \times 10^{-11}. 
\end{equation}
The uncertainty breakdown is presented in Table~\ref{tab4}, in the
same manner as it was done for $\overline{\mathcal B}^{\rm SM}_{s\mu}$
in Table~\ref{tab2}.

\begin{table}[t] 
\caption{The current uncertainty breakdown in $\overline{\mathcal B}_{d\mu}^{SM}$, as compared to the 2013 one. \label{tab4}}
\begin{tabular}{lccccccccc}\hline
& $f_{B_d}$ 
& CKM 
& $\tau_{\rm av}^d$ 
& $M_t$ 
& $\alpha_s$
& $\eta_{\scriptscriptstyle BBS}$
& other      
& non-        
& $\sum$
\\[-1mm]
&&&&&&&& parametric &\\\hline
2024 [this paper]
& $1.4$\%            
& $2.4\%$            
& $0.3$\%            
& $0.5$\%            
& $0.1$\%            
& $0.5$\%            
& $< 0.1$\%          
& $1.5$\%            
& $3.4$\%            
\\
2013~\cite{Bobeth:2013uxa} 
& $4.5$\%
& $6.9\%$ 
& $0.5$\%
& $1.6$\%
& $0.1$\%
& $0.0$\%
& $< 0.1$\%
& $1.5$\%
& $8.5$\%
\\\hline
\end{tabular}
\end{table}

\end{document}